\documentclass[pre,aps,twocolumn]{revtex4}
\usepackage{epsfig,psfrag}
\usepackage{bm}
\newcommand{\B}[1]{{\bm{#1}}} \newcommand{\C}[1]{{\mathcal{#1}}}
\newcommand{\Onecol} {\begin{widetext} \onecolumngrid} %% 2 -> 1
\newcommand{\Twocol} {\end{widetext} \twocolumngrid} %% 1 -> 2

\newcommand{\be}{\begin{equation}}
\newcommand{\ba}{\begin{array}} \newcommand{\bea}{\begin{eqnarray}}
\newcommand{\bfi}{\begin{figure}} \newcommand{\ee}{\end{equation}}
\newcommand{\ea}{\end{array}} \newcommand{\eea}{\end{eqnarray}}
\newcommand{\efi}{\end{figure}}

\begin{document}
\title{Eulerian Statistically Preserved Structures in Passive Scalar Advection}
  \author{Yoram Cohen$^{1,2}$ Anna Pomyalov$^1$ and
  Itamar Procaccia$^{1,2}$}
  \affiliation{$^1$ Dept. of Chemical Physics, The Weizmann Institute of
  Science, Rehovot 76100, Israel.\\
$^2$ Dept. of Physics, the Chinese University of Hong Kong, Shatin, Hong Kong.
} \pacs{}
\begin{abstract}
We analyze numerically the time-dependent linear operators that
govern the dynamics of Eulerian correlation functions  of a
decaying passive scalar advected by a stationary, forced
2-dimensional Navier-Stokes  turbulence.  We show how to naturally
discuss the dynamics in terms of effective compact operators that
display Eulerian Statistically Preserved Structures which
determine the anomalous scaling of the correlation functions. In
passing we point out a bonus of the present approach, in
providing analytic predictions for the time-dependent correlation
functions in decaying turbulent transport.
\end{abstract}
\maketitle

\section{Introduction}
The aim of this paper is to discuss the statistical physics of turbulent advection
of passive scalars \cite{79MY}. We are interested in scalar fields $\theta(\B r,t)$ which are
advected by a velocity field $\B u(\B r,t)$ such that together they solve
the set of equations
\begin{eqnarray}
\frac{\partial \B u}{\partial t}&+&(\B u\cdot \B \nabla)\B u=
-\B \nabla p +\nu \Delta \B u+\B f\ , \label{NS}\\
\frac{\partial \theta}{\partial t} &+&(\B u\cdot \B \nabla)\theta=
 \kappa \Delta \theta+f \ .\label{PS}
\end{eqnarray}
In these equations $p(\B r,t)$, $\nu$ and $\kappa$ are the pressure field, the kinematic viscosity
and the scalar diffusivity respectively. In this paper
the Navier-Stokes equations will be always forced
with a time-stationary forcing $\B f(\B r,t)$.
The scalar field will be forced or unforced (decaying), depending on our interests below.
We focus on the case of high Reynolds number Re and high Peclet number Pe, i.e.
$\nu, \kappa \to 0$, where the turbulence of the velocity field is fully developed and where
both fields display a significant range of scaling behavior at scales that are
sufficiently far from the forcing scales.

A major theoretical question that had been answered recently has to do
with the scaling properties of the correlation functions of the advected field \cite{98BGK,01CV}.
Define the simultaneous many-point correlation function $F^{(N)}(\B r_1, \dots \B r_N)$
in the forced case by
\begin{equation}
F^{(N)}(\B r_1, \dots \B r_N) \equiv \langle \theta(\B
r_1,t)\theta(\B r_2,t)\cdots \theta(\B r_N,t)\rangle_f \ ,
\label{defFn}
\end{equation}
with $\langle \cdots \rangle_f$ denoting an average with respect to realizations of the
stationary forcing $f(\B r,t)$ and of the velocity field $\B u (\B r,t)$.
It had been known for a long time that for high Re and Pe these functions
are homogeneous functions of their arguments, i.e.
\begin{equation}
F^{(N)}(\lambda \B r_1, \dots \lambda \B r_N) =\lambda^{\zeta_N} F^{(N)}(\B r_1, \dots \B r_N) \ ,
\end{equation}
with $\zeta_N$ being a scaling exponent that in general is {\em anomalous}, i.e. cannot be
guessed from dimensional considerations. But only recently it became clear how
these exponents are determined by the dynamical processes.

To understand the progress made, we rewrite the dynamical
equation for $\theta (\B r,t)$ in the shorthand form
\begin{equation}
\frac{\partial \theta(\B r,t)}{\partial t}=\C L \theta(\B
r,t)+f(\B r,t) \ , \label{defL}
\end{equation}
where in the present case $\C L\equiv \B u\cdot \B \nabla-\kappa
\Delta$.  In recent work \cite{01ABCPV,02CGP} it was clarified why
and how passive fields exhibit anomalous scaling, when the
velocity field is a generic turbulent field. The key is to
consider a {\em decaying problem} associated with Eq.
(\ref{defL}), in which the forcing $f(\B r,t)$ is put to zero.
The problem becomes then a linear initial value problem,
\begin{equation}
\partial \theta/\partial t =\C L \theta\ , \label{decay}
\end{equation}
with a formal solution
\begin{equation}
\theta(\B r,t) = \int d\B r' R(\B r,\B r',t) \theta(\B r',0) \
, \label{oper}
\end{equation}
with the operator
\begin{equation}
R(\B r, \B r', t)\equiv T^+ \exp[{\int_0^t ds \C L(s)}]\Big|_{\B r, \B r'}\ , \label{defR}
\end{equation}
and $T^+$ being the time ordering operator.
Define next the {\em time dependent} correlation
functions of the decaying problem:
\begin{equation}
C^{(N)}(\B r_1,\cdots ,\B r_N,t)\equiv \langle \theta(\B
r_1,t)\cdots \theta(\B r_N,t)\rangle \ . \label{defG}
\end{equation}
Here pointed brackets without subscript $f$ refer to the decaying
object in which averaging is taken with respect to realizations of
the velocity field and initial conditions. As a result of Eq. (\ref{oper}) the decaying
correlation functions are evolved by a propagator
$\C P^{(N)}(\underline{\B r},\underline{\B \rho},t)$, (with
$\underline {\B r}\equiv \B r_1,\B r_2,\!\cdots\!,\B r_N$ and $\underline {\B \rho}\equiv \B \rho_1,\B \rho_2,\!\cdots\!,\B \rho_N$)~:
\begin{equation}
C^{(N)}(\underline{\B r},t)=\int\!\! d\underline{\B \rho}
 \C P^{(N)}(\underline{\B r},\underline{\B \rho},t)~
C^{(m)}(\underline{\B \rho},0) \ . \label{defprop}
\end{equation}

In writing this equation we made explicit use of the fact that
the {\em initial} distribution of the passive field $\theta(\B r,0)$
is statistically independent of the advecting velocity field. Thus
the operator $\C P^{(N)}(\underline{\B r},\underline{\B \rho},t)$
can be written explicitly
\begin{equation}
\C P^{(N)}(\underline{\B r},\underline{\B \rho},t)\equiv
\langle  R(\B r_1,\B \rho_1,t)  R(\B r_2,\B \rho_2,t)\cdots
R(\B r_N,\B \rho_N,t)\rangle \ .
\end{equation}

The key finding \cite{01ABCPV,02CGP} is that the operator $\C
P^{(N)}(\underline{\B r},\underline{\B \rho},t)$ possesses a {\em
left} eigenfunction of eigenvalue 1, i.~e. there exists (for each
$N$) a time-independent function $Z^{(N)}(\underline{\B r})$ satisfying
\begin{equation}
Z^{(N)}(\underline{\B r})=\int d\underline {\B \rho}
Z^{(N)}(\underline{\B \rho}) \C P^{(N)}
(\underline{\B r},\underline{\B \rho},t)\ .\label{zeromode}
\end{equation}
The functions $Z^{(N)}$ are referred to as ``Statistically Preserved
Structures'', being invariant to the dynamics, even though
{\em the operator is strongly time dependent and decaying}. How to form,
from these functions, infinitely many conserved variables in the decaying
problem was shown in \cite{01ABCPV}, and is discussed again in
Sect. III. The functions
$Z^{(N)}(\underline{\B r})$ are homogeneous functions of
their arguments, with anomalous scaling exponents $\zeta_N$:
\begin{equation}
Z^{(N)}(\lambda\underline{ \B r}) = \lambda^{\zeta_N}
Z^{(N)}(\underline{\B r})+\dots
\end{equation}
where ``$\dots$'' stand for sub-leading scaling terms. Since
Eq.~(\ref{zeromode}) contains $Z^{(N)}(\underline{\B r})$ on both
sides, the scaling exponent $\zeta_N$ cannot be determined from
dimensional considerations, and it can be anomalous. More
importantly, it was shown that the correlation functions of the
forced case, $F^{(N)}(\underline{\B r})$ Eq. (\ref{defFn}), have
exactly the same scaling exponents as $Z^{(N)}(\underline{\B r})$
\cite{02CGP}. In the scaling sense
\begin{equation}
F^{(N)}(\underline{\B r})\sim Z^{(N)}(\underline{\B r}) \ .
\label{scalesame}
\end{equation}
This is how anomalous scaling in passive fields is understood.

Besides exactly soluble examples in which the advecting velocity
field is non-generic (i.e. $\delta$-correlated in time) the
existence of eigenfunctions of eigenvalue 1 of the time dependent
propagator was demonstrated fully only in shell models of
turbulence. While the present authors believe that shell models
contain a lot of the robust properties of real turbulence, this
belief is not universally accepted in the community. It is
therefore necessary to demonstrate that the mechanism sketched
above exists indeed in the full problem Eqs.
(\ref{NS})-(\ref{PS}). This had been done for 3'rd order
correlations within the Lagrangian formulation in \cite{01CV}.
The aim of this paper is to demonstrate this in the Eulerian
frame, and for correlation functions of order 2, 4 and 6.

In Section II we describe the simulations of Eqs. (\ref{NS}) and
(\ref{PS}) and the type of measurements that we performed. Sect.
III is a theoretical digression, in which we analyze an exactly
soluble simple model to guide ourselves as to how to analyze the
numerical results to find the scaling forms of the $n$th oder
propagators and their eigenfunctions. Sect. IV describes the
analysis of the data, and Sect. V offers a discussion and a
summary of the paper.
%%%%%%%%%%%%%%%%%%%%%%%%%%%%%%%%%%%%%%%%%%%%%%%%%%%%%%%%%%%%%%%%%%%%%%%
\section{Simulations}
We performed a Direct Numerical Simulation (DNS) of Eqs.
(\ref{NS}) and (\ref{PS}) on a $2048\times 2048$ 2-D array. The
forcing $\B f$ in Eq. (\ref{NS}) is random, $\delta$ correlated
in time, isotropic and homogeneous. Its $k$ depdendence is
\begin{equation}
\langle|\B f(\B k)|\rangle \propto k\;exp[-0.5(k/1024)^2]\ .
\end{equation}
This forcing is biased towards the small scales; this is done
because of the inverse energy cascade that characterizes
2-dimensional turbulence. The fluid dissipation is modeled by a
hyper-viscosity  term proportional to $\Delta^8 \B u$. In
addition we employed  a ``friction'' term proportional to $\B u$
in order to stabilize the velocity field on the largest scales.
The passive field $\theta$ dissipates normally as shown in Eq.
(\ref{PS}).

The simulations were performed for a decaying passive field
$\theta$, that is, the forcing $f$ in Eq. (\ref{PS}) was put to
zero. The initial conditions for the $\theta$ field were of the
form
\begin{equation}\label{init}
\tilde \theta(\B k,t=0)=\delta(k-k_0)exp(i\gamma(\B k))\, ,
\end{equation}
where $\tilde \theta(\B k,t)$ is the Fourier transform of the
real space variable defined as:
\begin{equation}
\tilde h(\B k)=\frac{1}{2\pi}\int d \B r h(\B r) e^{-i \B k \B r}
\end{equation}
(from now on we will omit the tilde above the functions and denote the k
space functions only by their variables).
$\gamma(\B k)$ is a random variable in the interval
$\gamma\in [0,2 \pi]$ where $\gamma(\B k)=-\gamma(-\B k)$, insuring that
$\theta^*(-\B k,0)=\theta(\B k,0)$, and therefore that $\theta(\B
r,t)$ is real.

As the initial conditions for $\theta$ and the forcing of the $\B
u$ fields are both homogeneous and isotropic so are the scalar
correlation functions $C^{(N)}(\B r_1 \dots \B r_N,t)$ defined in
Eq. (\ref{defG}), at all times $t$. We measured the $\B k$ space
correlation functions:
\begin{eqnarray}
&&C^{(N)}(\B k_1,\dots,\B k_{N},t)\delta(\B k_1+\dots+\B
k_{N})\nonumber\\ &&=\langle \theta(\B k_1,t)\dots \theta(\B
k_N,t)\rangle \ , \label{defCm}
\end{eqnarray}
where again the average $\langle \dots\rangle$ is over initial
conditions and over realizations of the $\B u$ field. The delta
function appears due to translational invariance.

In accordance to Eq. (\ref{defprop}) we define the $\B k$ space propagator:
 \begin{equation}
C^{(N)}(\underline{\B k},t)=\int\!\! d\underline{\B k}'
 \C P^{(N)}(\underline{\B k},\underline{\B k}',t)~
C^{(N)}(\underline{\B k}',0) \ . \label{defpropk}
\end{equation}

\subsection{The 2-point propagator}
The only two point correlator that is not zero is $C^{(2)}(\B
k,-\B k,t)=\langle\theta(\B k)\theta^*(\B k)\rangle$. Because of
the isotropy of the initial conditions and the driving field, the
correlator depends only on the magnitude of $\B k$. We can
therefore consider the second order structure function
$S^{(2)}(k,t)=C^{(2)}(\B k,-\B k,t)$, and it's propagator $\hat \C
P^{(2)}(k,k',t)$.

\begin{figure}
\centering
\psfrag{yl}{$\hat\C P^{(2)}(k,300,t)$}
\psfrag{xl}{$k$}
\includegraphics[width=.45\textwidth]{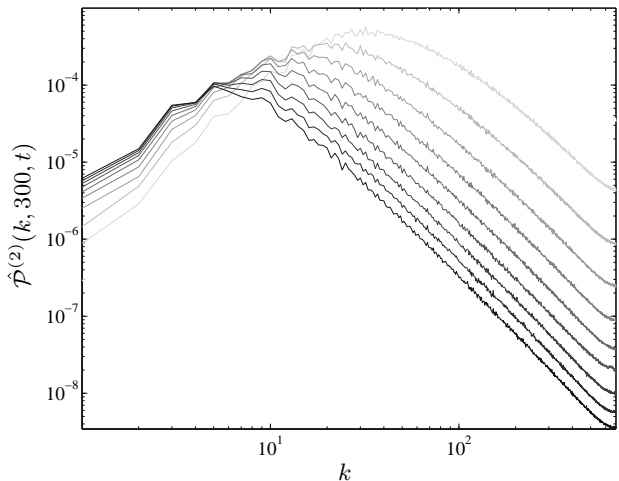}
\caption{$\hat \C P^{(2)}(k,300,t)$ for 10 different times. The
first time is $\tau_{30}$, i.e. the eddy turn over time at scale
$k=30$, in lightest grey. Later times, in darker and darker greys,
are at 2$\tau_{30}$, 3$\tau_{30}$, etc. until 10$\tau_{30}$.}
\label{fig1}
\end{figure}

In discrete $k$-space (as in our simulation on a grid) the
propagator $\hat \C P^{(2)}(k,k',t)$ has a matrix representation.
For the choice of initial conditions as in Eq. (\ref{init}),
$S^{(2)}(k,t)$ is simply the $k_0$th column of the
propagator :
\begin{equation}
S^{(2)}(k,t)=\int\!\! dk'
 \hat \C P^{(2)}(k,k',t)~
S^{(2)}(k',0) =\hat \C P^{(2)}(k,k_0,t)\ . \label{defpropd2}
\end{equation}
In Figure \ref{fig1} we plot such a column of the 2-point
propagator, at 10 different times. Two properties of the
propagator should be noticed: as time progresses the overall
amplitude decreases due to the decay, while the maximum moves
from the initial $k_0$ to lower values of $k$. In Sects. III and
IV we will find the scaling form of this propagator.
%%%%%%%%%%%%%%%%%%%%%%%%%%%%%
%%%%%%%%%%%%%%%%%%%%%%%%
\subsection{Multi-point propagators}
In the case of the multi-point correlator the overall $\delta$
function and the isotropy of the fields do not reduce the
dependence to a one variable function. The propagators are
therefore functions of many $k$-vectors. It turns out however that
measurement of the statistics for objects depending on many
$k$-vectors are very taxing. We opted therefore the extract from
the DNS partial information on the dynamics of the $2N$ order
structure functions with $N>1$:
\begin{equation}\label{gfus}
S^{(2N)}(k,t)=\langle|\theta(\B k,t)|^{2N}\rangle \ .
\end{equation}
Accordingly we define the reduced propagators :
\begin{eqnarray}
S^{(2N)}(k,t)&=&\int\!\! dk'
 \hat \C P^{(2N)}(k,k',t)~
S^{(2N)}(k',0) \ .
\label{defpropdmult}
\end{eqnarray}

\begin{figure}
\centering
\psfrag{yl}{$\hat\C P^{(4)}(k,300,t)$}
\psfrag{xl}{$k$}
\includegraphics[width=.45\textwidth]{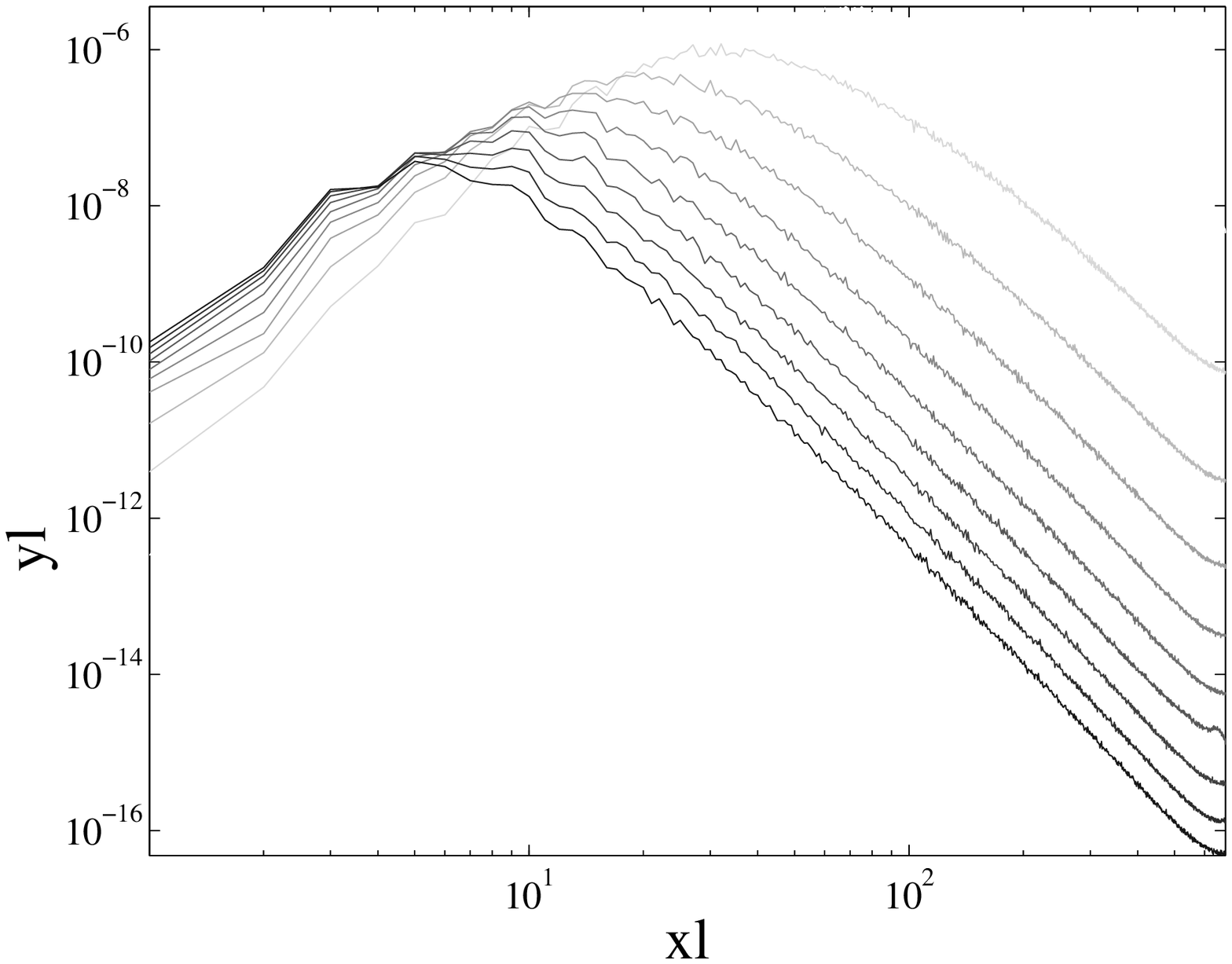}
\psfrag{yl}{$\hat\C P^{(6)}(k,300,t)$}
\psfrag{xl}{$k$}
\includegraphics[width=.45\textwidth]{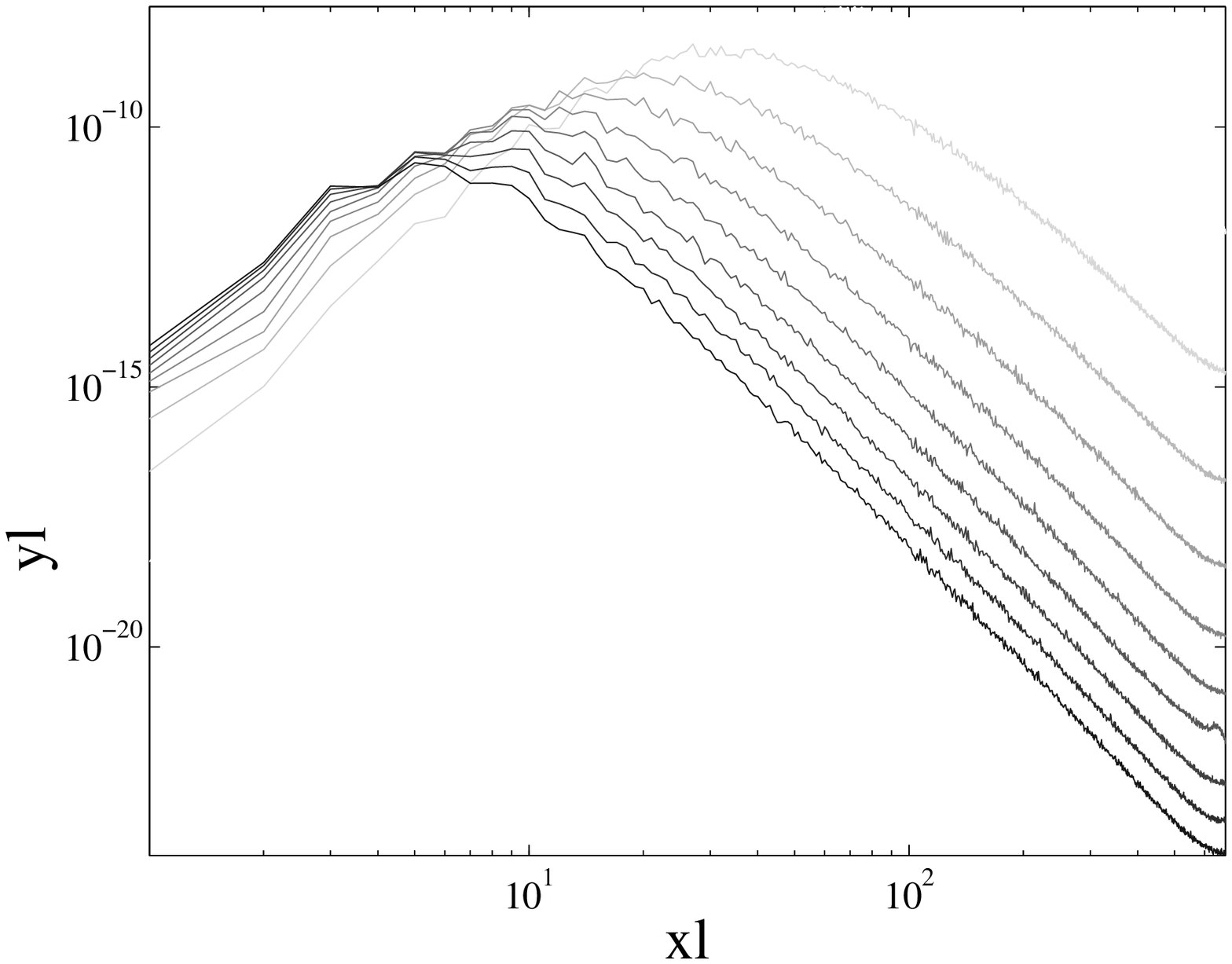}
\caption{$\hat \C P^{(4)}(k,300,t)$ (upper panel) and $\hat \C
  P^{(6)}(k,300,t)$ (lower panel) for the same 10 different
times as in Fig. \ref{fig1}} \label{fig2}
\end{figure}

In Figure \ref{fig2} we present the numerical
results for of the fourth and sixth order reduced propagators, at
the same ten different times. These objects display similar
qualitative behavior to that of the 2-point propagator. We will
learn in the next section how to think about the scaling
properties of these objects and how to re-plot the numerical
results in proper rescaled variables.
%%%%%%%%%%%%%%%%%%%%%%%%%%%%%
%%%%%%%%%%%%%%%%%%%%%%%%%%%%%%%%%%%%%%%%%%%%%%%%%%%%%%%%%%%%%%%%
\section{An Exactly solvable case}
In order to motivate the data analysis presented in Sect. IV we
turn now to the Kraichnan model for passive scalar advection. The
model is exactly solvable, and examining the analytic forms of the
propagators offers clues to what should be expected in the
generic case. In principle we could perform this analysis in
terms of the Kraichnan model for a 2-dimensional passive scalar.
In fact, it is sufficient to consider the shell-model version;
the latter is very transparent to analytic manipulations, and for
our purposes the results throw equally useful light on how to
perform the analysis of the DNS results.

\subsection{The Kraichnan shell model}
The Kraichnan shell model for passive scalar advection \cite{97BBW,01ABCPV}, as all
other shell models of turbulent flows, is written in terms of
Fourier components of the field. The Fourier components are
restricted to shells denoted by the index $m$, and the equation
takes on the form
\begin{eqnarray}
{d\theta_m\over dt}&=&\C L_{m,n}\theta_n \label{ldef}\\
\C L_{m,n}&=&ik_{m+1}u_{m+1}\delta_{m+1,n}
+ik_m u^*_{m}\delta_{m-1,n}-\kappa k_m^2\delta_{m,n}\ .\nonumber
\label{pshell}
\end{eqnarray}
Here $k_m$ are the shell $k$ vector, $k_m=k_0 \lambda^m$ for some
$k_0$ and $\lambda$. The shell components of the velocity field,
$u_m(t)$, are Gaussian random variables, $\delta$-correlated in time,
satisfying:
\begin{eqnarray}\label{Vcort}
\langle u_n(t)u^*_m(t')
\rangle=c_0~\delta_{n,m}\delta(t-t')~\lambda^{-\xi m} \ .
\end{eqnarray}
Here $\xi$ is the scaling exponent of the $u$ field.
%%%%%%%%%%%%%%%%%%%%%%%%%%%%%%%%%%%%%%%%%%%%%%%%%%%%
\subsection{The 2nd order propagator}
The second order correlator satisfies the explicit equation (see
\cite{02CGP}):
\begin{equation}\label{kri}
\frac{d}{dt}S^{(2)}_n(t) =\Big(M^{(2)}_{n,m} -\kappa
k_n^2\Big)S^{(2)}_m(t) \, ,
\end{equation}
$S^{(2)}_n(t)$ is the shell equivalent of the 2nd order structure
function in Eq. (\ref{defpropd2}),
\begin{equation}\label{kri1}
S^{(2)}_n(t)=\langle|\theta_n(t)|^2\rangle \ .
\end{equation}
The time evolution operator $\B M^{(2)}$ has the form
\begin{eqnarray}\label{krim1}
M^{(2)}_{n,m}&=& (\alpha_n+\alpha_{n+1})\delta_{n,m}
\\&-&\alpha_n\delta_{n-1,m}
-\alpha_{n+1}\delta_{n+1,m}\ ,\nonumber
\end{eqnarray}
where
\begin{equation}
\alpha_n\equiv -c_0k_0^2 \left(\frac{k_n}{k_0}\right)^{\zeta_2}
=-c_0k_0^2 \lambda^{\zeta_2 n} \ , \label{alphalam}
\end{equation}
and $\zeta_2=2-\xi$ is the dimensional scaling of the 2-point
correlation function. The operator $\B M^{(2)}$ has the following useful
scaling property:
\begin{equation}
\lambda^{-\zeta_2 p} M_{n+p,m+p}^{(2)}=M_{n,m}^{(2)} \ .
\label{propM}
\end{equation}
The second order propagator has, in the limit of vanishing
viscosity, the explicit form:
\begin{equation}\label{krip}
\C P^{(2)}_{n|m}(t)=\exp(t\B M^{(2)})\Big|_{n,m}\ .
\end{equation}

Note that the time evolution operator and the propagator are both
Hermitian, and thus admit an eigenvector decomposition. The time
evolution operator has two types of eigen-vectors $\psi^{(2,q)}_n$ and
$\tilde \psi^{(2,q)}_n$
which we can regard as slow modes and fast modes respectively. Here
the index $k$ stands for the eigen-mode index. The fast modes are
dominated by the viscous term, their support is in the viscous range,
and they can essentially be taken to be $\tilde \psi^{(2,q)}_n=\delta_{n,q}$
for some shell $q$  with the shell index $q > m_d$, above which viscosity
dominates Eq. (\ref{kri}). The transition shell $m_d$ is determined by the
condition $\kappa k^2_{m_{d}}= c_0k^{\zeta_2}_{m_{d}}$.
For the fast modes we have the following approximate Equation:
\begin{equation}
\frac{d}{dt}\tilde \psi^{(2,q)}_n = -\kappa k_n^2\tilde \psi^{(2,q)}_n \, ,
\end{equation}
therefore their eigen-values are $\beta_q =-\kappa k_q^2$.

For slow modes, $\psi^{(2,q)}_n$,  which have their support on
shells smaller than $m_d$, the dissipative term can be neglected
and they satisfy an eigen-function equation of the form:
\begin{equation}
\beta_q \psi^{(2,q)}_n=M^{(2)}_{n,m}\psi^{(2,q)}_m \ .
\end{equation}

For a sufficiently large inertial range we can use the scaling
property (\ref{propM}) to get:
\begin{equation}
\beta_q \psi^{(2,q)}_n=\lambda^{-\zeta_2p}
M^{(2)}_{n+p,m+p}\psi^{(2,q)}_{m} \ .
\end{equation}
Shifting indices we rewrite,
\begin{equation}
\beta_q \psi^{(2,q)}_{n-p}=\lambda^{-\zeta_2p}
M^{(2)}_{n,m}\psi^{(2,q)}_{m-p}\ .
\end{equation}
Define now a vector
\begin{eqnarray}\label{sl1}
\chi_{n}\equiv\psi^{(2,q)}_{n-p} \, ,
\end{eqnarray}
we have:
\begin{eqnarray}
\lambda^{\zeta_2 p}\beta_q \chi_{n}&=&M^{(2)}_{n,m}\chi_{m} \ ,
\end{eqnarray}
We can thus define $\psi^{(2,q+p)}_{n}\equiv \chi_n$,  an
eigen-function of the time evolution operator $\B M^{(2)}$ with
eigen-value $\lambda^{\zeta_2 p} \beta_q$.

We can thefore conclude that the eigen-vectors and eigen-values may
be obtained from each other by shift of the indices:
\begin{eqnarray}
\psi^{(2,q+p)}_{n+p}&=&\psi^{(2,q)}_{n} \label{psi_shift}\ ,\\
\beta_{q+p}&=&\lambda^{\zeta_2 p}\beta_{q}  \label{beta_shift}\ . 
\end{eqnarray}

\begin{figure}
\centering
\psfrag{xl}{$n$}
\psfrag{yl}{$\psi^{(2,15)}_n,\,\psi^{(2,35)}_{n+20}$}
\includegraphics[width=.5\textwidth]{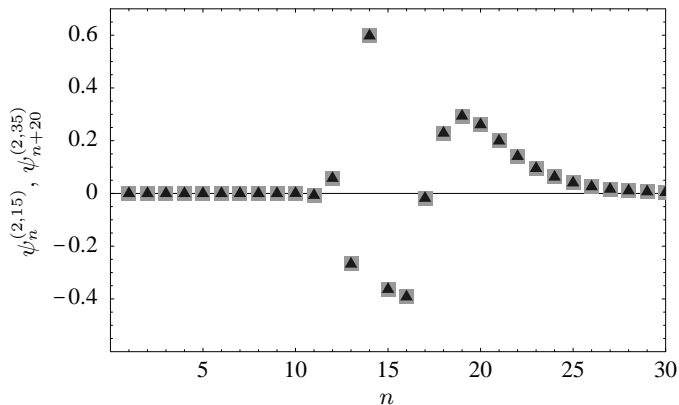}
\caption{A plot of the eigen-vectors
 $\psi^{2,15}_n$ (squares), and $\psi^{2,35}_{n+20}$ (triangles),
 for $n\in[0,30]$  }
\label{multi1}
\end{figure}

Fig.~\ref{multi1} demonstrates that indeed two different
eigen-functions coincide once shifted with respect to each other.
It follows from (\ref{beta_shift}) that
\begin{eqnarray}\label{2pa}
\beta_{q}\propto \alpha_q.
\end{eqnarray}

\begin{figure}
\centering
\psfrag{xl}{$n$}
\psfrag{yl}{$\alpha_n$}
\includegraphics[width=.5\textwidth]{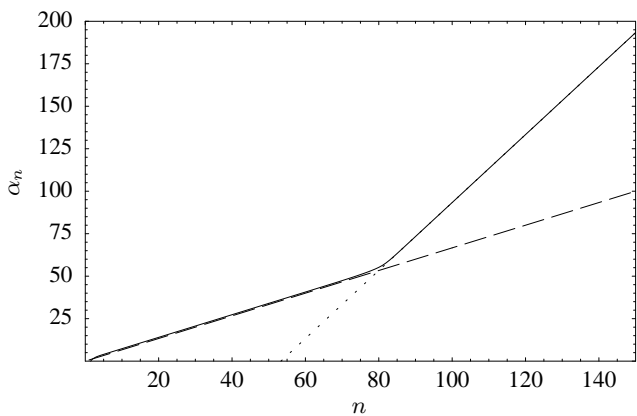}

\caption{The eigen-values, plotted from small to large. The
eigen-values are well fitted by the analytic predictions
$k_m^{\zeta_2}$ (dashed line) for the slow modes, and by
$k_m^{2}$ (dotted line) for the fast modes. The transition to viscous
range occur at shell $m_d=80$ }
\label{multi2}
\end{figure}

In Fig.~\ref{multi2} the spectrum of the eigen-values is plotted,
showing the two expected regions: a region for which the scaling
of the eigen-values is $\propto k_n^{\zeta_2}$  for the slow
modes, and  $\propto k_n^{2}$ for the fast modes.
%%%%%%%%%%%%%%%%%%%%%%%%%%%%%%%%%%%%
%%%%%%%%%%%%%%%%%%%%%%%%%%%%%%%%%%%%%%
%%%%%%%%%%%%%%%%%%%%%%%%%%%%%%%%%%%%
%%%%%%%%%%%%%%%%%%%%%%%%%%%%%%%%%%%%%%
The lowest eigenvalue is proportional to $\alpha_0$. The
eigen-function $\psi^{(2,0)}_n$ associated with it can be calculated
explicitly, and it exhibits a normal, dimensional, scaling
in the inertial range:
\begin{eqnarray}\label{zmod}
\psi^{(2,0)}_n \propto (k_n/k_0)^{-\zeta_2} \ .
\end{eqnarray}
Using  Eqs .(\ref{psi_shift}) and (\ref{zmod}), that 
imply that $\psi^{(2,q)}_n$ has a scailing 'tail' with exponent $\zeta_2$, 
starting after shell $q$ we establish:
\begin{eqnarray}
\psi^{(2,q)}_n\propto
(k_n/k_q)^{-\zeta_2}\propto\alpha_q/\alpha_n\, \quad q<n<m_d \ ,
\end{eqnarray}

We now use these results to learn how to re-scale our numerical
data. Suppose that we started with an initial condition in some
shell $n$ in the inertial range, $S^{(2)}_p(0)=\delta_{p,n}$. The
fast modes do not contribute significantly to the time
dependence, since they decay fast and anyhow have no support in
the inertial range. Thus, using only the slow modes and Eqs.
(\ref{krim1}),(\ref{krip}) , we have for $p<n$,
\begin{eqnarray} \label{phi1}
S^{(2)}_p(t)&=&\sum_{q=1}^{p}\psi^{(2,q)}_n
e^{\alpha_qt}\psi^{(2,q)}_p\nonumber\\&\propto& \frac{1}{\alpha_p
\alpha_n}\sum_{q=1}^{p}\alpha_q^2e^{\alpha_qt}\propto {\cal
P}^{(2)}_{p|n}(t) \ .
\end{eqnarray}
In going from the first to the second line we inserted the
$\delta$-function initial conditions, getting thus a column of the
propagator ${\cal P}^{(2)}_{p|n}(t)$. Remember that in our DNS
calculations we fixed the column index, and are interested in the
scaling behavior with respect to the row index and time. Therefore
we compute now (up to an overall dimensional constant),
\begin{eqnarray}\label{phi2}
&&\lambda^{\zeta_2 m} {\cal P}^{(2)}_{p-m|n}(\lambda^{\zeta_2 m}
t)= \frac{\lambda^{\zeta_2 m}
}{\alpha_{p-m}\alpha_n}\sum_{q=1}^{p-m}\alpha_q^2e^{\alpha_{q+m}t}\nonumber\\
&&=\frac{1}{\alpha_p\alpha_n}\sum_{q=1}^{p-m}\alpha_{q+m}^2e^{\alpha_{q+m}t}
=\frac{1}{\alpha_p\alpha_n}\sum_{q=m+1}^{p}\alpha_{q}^2e^{\alpha_{q}t}\nonumber
\\
&&= {\cal P}^{(2)}_{p|n}(t)-\frac{\alpha_m}{\alpha_p}{\cal
P}^{(2)}_{m|n}(t) \ ,
\end{eqnarray}
where we have used the explicit form of the propagator in Eq. (\ref{phi1})
and Eq. (\ref{alphalam}). For $m$ much smaller
than $p$ we can neglect the second term. Then we conclude that
the propagator is a homogeneous function of the variables $k_p$
and $t$. Explicitly, multiplying by $t$,
\begin{equation}\label{prp2}
t{\cal P}^{(2)}(k_p,k_n,t)=(\lambda^{-\zeta_2 m}t){\cal
P}^{(2)}(\lambda^m k_p,k_n,\lambda^{-\zeta_2 m} t)
\end{equation}
For fixed $k_n$ this is a homogeneous function of two arguments
which can be always written in the form
\begin{equation}
{\cal P}^{(2)}(k_p,k_n,t) = \frac{\text{Const}}{\alpha_n t}
\Lambda(\alpha_p t) \ ,
\end{equation}
for some function $\Lambda(x)$. The symmetry between $p$ and $n$
is restored by realizing that the asymptotic form of $\Lambda(x)$ is
$1/x$ for large $x$.

This scaling form was also found in the study of the generic shell
models of passive scalar advection (arbitrary time dependence in the
velocity correlations) \cite{02CGP} and we will show that the same scaling form
survives when we go back to the generic model studied by DNS.
%%%%%%%%%%%%%%%%%%%%%%%%%%%%%%%%%%%%%%%%%%%%%%%%%%%%%%%%%%%%%%%%%%%%
\subsection{The multi point propagators}
The invariance of Eq. (\ref{pshell}) under a uniform phase change
$\theta_n\rightarrow e^{i\phi}\theta_n$ dictates that the only
non-zero correlation functions will have an equal number of
variables $\theta_n$ and conjugated variables $\theta^*_m$. We
define:
\begin{equation}\label{gkmult}
C^{(2N)}_{i_1,\dots,i_m,j_1,\dots,j_N}(t) =\langle
\theta_{i_1}(t)\dots \theta_{i_N}(t)\theta^*_{j_1}(t)\dots
\theta^*_{j_N}(t)\rangle
\end{equation}
As in the 2-point case we can define the respective multi-point
differential time derivative operator, analogous to Eq.
(\ref{kri}). In the limit of vanishing viscosity:
\begin{eqnarray}
\frac{d}{dt}C^{(2N)}_{\underline{\B i},\underline{\B
    j}}(t)=M^{(2N)}_{\underline{\B i},\underline{\B j}|\underline{\B
    i}',\underline{\B j}'}C^{(2N)}_{\underline{\B i}',\underline{\B
    j}'}(t)
\end{eqnarray}
The respective propagator is defined by:
\begin{eqnarray}
C^{(2N)}_{\underline{\B i},\underline{\B j}}(t) &=&\C
P^{(2N)}_{\underline{\B i},\underline{\B j}|\underline{\B
    i}',\underline{\B j}'}(t)
C^{(2N)}_{\underline{\B i}',\underline{\B j}'}(0)\label{prop2m}\\
\C P^{(2N)}_{\underline{\B i},\underline{\B j}|\underline{\B
    i'},\underline{\B j'}}&=&\exp(t \B M^{(2N)})\Big|_{\underline{\B
    i},\underline{\B j}|\underline{\B i}',\underline{\B j}'} \ .
\end{eqnarray}

Because of the time reversibility of the statistics of the $\B u$
fields, and the anti-hermiticity of the un-averaged differential
operator $\C L$ in Eq. (\ref{ldef}): $ \C L^*_{m,n}=-\C L_{n,m}$,
both the time derivative operator $\B M^{(2N)}$ and the
propagator itself are hermitian, and therefore admit eigenvalue
decomposition. Furthermore the operator $\B M^{(2N)}$ has the
following scaling property:
\begin{eqnarray}
M^{(2N)}_{\underline{\B i},\underline{\B j}|\underline{\B i'},\underline{\B
    j'}}=
\lambda^{-\zeta_2 p}M^{(2N)}_{\underline{\B i}+p,\underline{\B j}
+p|\underline{\B i'}+p,\underline{\B j'}+p} \ . \label{defM2m}
\end{eqnarray}

As in the two point case the dynamics within the scaling range is
determined by slow modes.
\begin{equation}\label{4eig}
\beta^{(2N,l)}_{k}\psi^{(2N,l,k)}_{\underline{\B i},\underline{\B
j }}=M^{(2N)}_{\underline{\B i},\underline{\B j}|\underline{\B
i}',\underline{\B j}'}\psi^{(2N,l,k)}_{\underline{\B
i}',\underline{\B j}'} \ .
\end{equation}
In this equation the index $l$ stands for a family of eigen-modes,
and $k$ for their index within the family. Each $l$ family can be
obtained from any one of its members by shifting the indices. The
eigen-values of the modes within a given family can be obtained
also by a shift :
\begin{eqnarray}
\psi^{(2N,l,k+p)}_{\underline{\B i}+p,\underline{\B
j}+p}&=&\psi^{(2N,l,k)}_{\underline{\B i}
,\underline{\B j}}\label{gamshift}\\
\beta^{(2N,l)}_{k+p}&=&\lambda^{\zeta_2 p}\beta^{(2N,l)}_{k} \ .
\label{scalesim}
\end{eqnarray}
Note that in the 2-point case we had only one family of
eigen-modes. Here we added the index $l$ to the eigen-values and
eigen-modes to distinguish the different families. Note also that in
Eq. (\ref{scalesim}) the scaling exponent of the eigenvalues is
$\zeta_2$; this stems from the scaling properties of the
differential operator $\B M^{(2N)}$, cf. eq. (\ref{defM2m}).
However, the eigen-modes display in general anomalous scaling
which can be represented by
\begin{eqnarray}
\psi^{(2N,l,k)}_{\underline{\B i},\underline{\B j}}
=\lambda^{-p~\zeta_{2N,l}}~ \psi^{(2N,l,k)}_{\underline{\B
i}+p,\underline{\B j}+p} \ , \label{shift}
\end{eqnarray}
where $\zeta_{2N,l}$ is the (anomalous) scaling exponent of the
$l$th family of eigen-modes.

Since the DNS were analyzed in terms of fused objects, we focus
on initial conditions in Eq.(\ref{prop2m}) for which all the $2N$
indices are the same. We also measure the resulting structure
functions
\begin{equation}
S^{(2N)}_n=\langle |\theta_n|^{2N}\rangle \ . \label{S2m}
\end{equation}
This procedure will extract a fused propagator for which there
are only two indices, and we denote it bellow as ${\cal
P}^{(2N)}_{p|n}(t)$. This propagator is not Hermitian, it has in
general no eigen-function decomposition, but we can understand the
scaling form of any of its columns from the knowledge of the full
propagator and its eigen-functions.

The equivalent of Eq. (\ref{phi1}) for
$S_k^{(2N)}(t=0)=\delta_{k,n}$, $p<n$, summing over all families
of slow modes (i.e. over the index $l$) is:
\begin{eqnarray} \label{phi1m}
&&S^{(2N)}_p(t)=\sum_{l}\sum_{q=1}^{p}\psi^{(2N,l,q)}_n
e^{\alpha_qt}\psi^{(2N,l,q)}_p\nonumber\\
&&=\sum_{l}\frac{C_l}{(\alpha_p
\alpha_n)^{\frac{\zeta_{2N,l}}{\zeta_2}}}
\sum_{q=1}^{p}\alpha_q^\frac{2\zeta_{2N,l}}{\zeta_2}e^{\alpha_qt}={\cal P}^{(2N)}_{p|n}(t) \ .
\end{eqnarray}
In going from first to second line we have used the fact that

\begin{equation}
\psi_{i,i,\dots,i}^{2N,l,p} \propto
\left(\frac{\alpha_i}{\alpha_p}\right)^{-\zeta_{2N,l}/\zeta_2} \ .
\end{equation}
This follows directly from Eqs. (\ref{gamshift}) and
(\ref{shift}).

Using the same argumentation as in Eq. (\ref{phi2}) we get:
\begin{eqnarray}\label{phimult}
\C P^{(2N)}_{p|n}(t)=\sum_l \frac{C_l}{(\alpha_nt)^\frac{\zeta_{2N,l}}{\zeta_2}}
\Lambda^{(2N,l)}(\alpha_pt) \, ,
\end{eqnarray}

here again the $C_l$'s are dimensional constants and the functions
$\Lambda^{(2N,l)}(x)$ are some functions with the asymptotic form
$\Lambda^{(2N,l)}(x)\approx x^{-\frac{\zeta_{2N,l}}{\zeta_2}}$ for $x$
large. We note that for
sufficiently long times and if the scaling exponents are well
separated, we can expect the sum to be dominated by the leading
scaling exponent.
\section{Data Analysis}
The detailed analysis that was possible for the simple model of
Sect. III is not available for the generic model Eqs.(\ref{NS}),
(\ref{PS}). Our strategy is to assume that the scaling forms
derived in the last section are still valid for the generic
model, and to demonstrate this by replotting the data
accordingly. We will see that these predictions are born out by
the data.
%%%%%%%%%%%%%%%%%%%%%%%%%%%%%%%%%%%%%%%%%%
\subsection{analysis of the 2-point porpagator}
\begin{figure}
\centering
\psfrag{yl}{$t \hat\C P^{(2)}(k,300,t)$}
\psfrag{xl}{$k t^{1/\zeta_2}$}
\includegraphics[width=.45\textwidth]{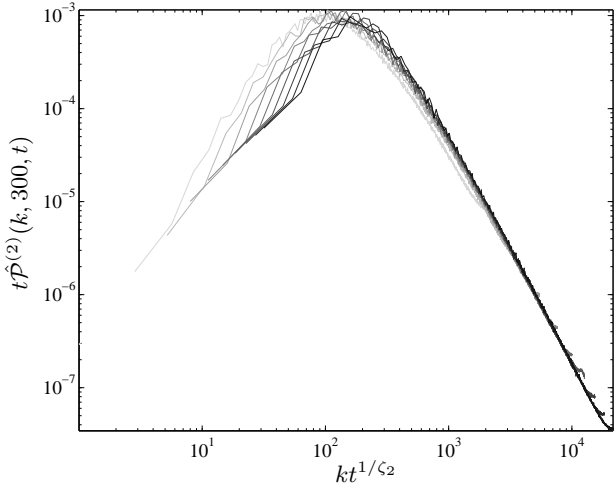}
\caption{$t\hat \C P^{(2)}(k,300,t)$ for 10 different times
(light grey earliest dark grey last)} \label{fig4-1}
\end{figure}

In light of  Eq. (\ref{prp2}), we expect the 2nd order propagator
$\hat \C P^{(2)}(k,k_0,t)$ to be, for a fixed $k_0$, a
homogeneous function of the variable $kt^{1/\zeta_2}$, and to
decay as $1/t$:
\begin{equation}
\hat \C
P^{(2)}(k,k_0,t)\propto\frac{1}{t}H^{(2)}\Big(kt^{1/\zeta_2}\Big)
\ ,
\end{equation}
where $H^{(2)}$ is some function. We test the correctness of this
form in Fig. \ref{fig4-1}. To this aim we re-plot that data shown
in Fig. \ref{fig1} in different coordinates, multiplied by $t$ and
as a function of
\begin{equation} \hat k=kt^{1/\zeta_2}\ . \end{equation}
The quality of the data collapse appears to
strongly support the proposed scaling form. We note that the data
collapse is superior on the right of the maximum, and less
convincing at its left. We believe that this stems from two
reasons. First, there is better statistics for the right part of
the curve, simply because it belongs to larger $k$ vectors where
the angular average is more extensive. Second, the left part of
the curve is more sensitive to the finite size effects,
particularly to the fact that the driving velocity field loses its
scaling form close to $L$. With all the limitations of the
numerical simulations we consider the data collapse as very
satisfactory.
%%%%%%%%%%%%%%%%%%%%%%%%%%%%%
%%%%%%%%%%%%%%%%%%%%%%%%

\begin{figure}
\centering
\psfrag{yl}{$\int_0^\infty \hat \C P^{(2)}(k,300,t)dk$}
\psfrag{xl}{$t$}
\includegraphics[width=.45\textwidth]{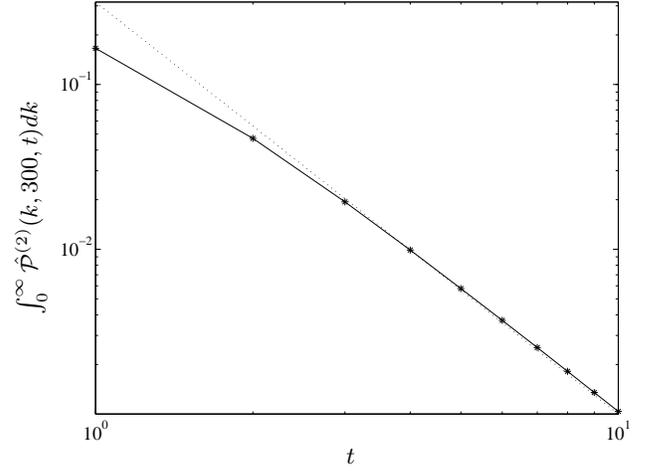}
\caption{The integral of the 2-point propagator, $\int_0^\infty \hat \C
  P^{(2)}(k,300,t)d^2k$, (solid line) as a function of time. The dotted line
  is the expected $t^{-(1+1/\zeta_2)}$. (time is
  measured in this figure, and in the ones that follow in units of $\tau_{30}$ the eddy turnover
  time of $k=30$)} \label{fig4-2}
\end{figure}

If the prediction for the 2-point function holds, then we will have for the
time dependent integral:
\begin{equation}\label{t2}
\int_0^\infty \hat \C P^{(2)}(k,k',t)dk \approx
\frac{1}{t^{1+1/\zeta_2}}\int_0^\infty  H^{(2)}(\hat
k )d\hat k\propto\frac{1}{t^{1+1/\zeta_2}}
\end{equation}
In Fig. \ref{fig4-2}, We show that indeed after an initial period
the integral settles on a scaling form consistent with Eq.
(\ref{t2}).
%%%%%%%%%%%%%%%%%%%%%%%%%%%%%
%%%%%%%%%%%%%%%%%%%%%%%%

\begin{figure}
\centering
\psfrag{yl}{$\int Z^{(2)}(k)\C P^{(2)}(k,300,t) d^2 \B k$}
\psfrag{xl}{$t$}
\includegraphics[width=.45\textwidth]{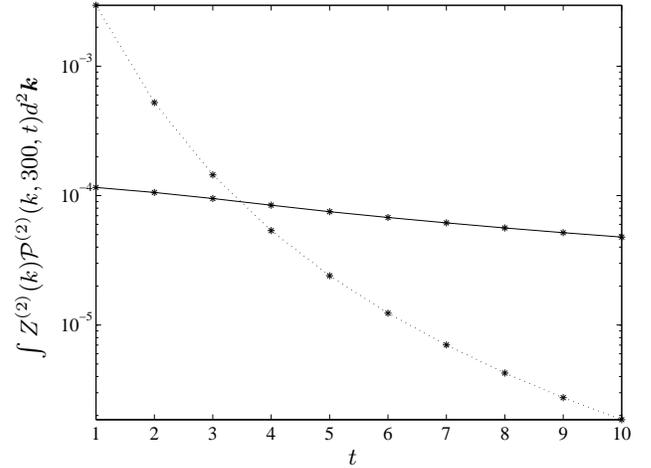}
\caption{The integral of the 2-point propagator $P^{(2)}(k,300,t)$ weighed by the left zero mode
$Z^{(2)}(k)$ (solid line),
  compared to the integral of the un-weighed second order structure function in dotted line.}
\label{fig4-3}

\end{figure}

Using the form of the propagator we can establish the existence
of a left zero mode of the 2-point propagator. Integrating over
the two sides of Eq. (\ref{zeromode}) we have a time independent
expression on the left hand side. We therefore expect that the
weighed integral of the propagator with the function $Z^{(2)}$
will be constant:
\begin{eqnarray}
&&\int  Z^{(2)}(\B k)\C P^{(2)}(\B k,\B k',t) d^2 \B k\\
&&\propto\int_0^\infty  k^{-2-\zeta_2}\frac{H^{(2)}(k t^{1/\zeta_2})}{t}k dk=\int_0^\infty \frac{H(\hat{k})}{\hat k^{1+\zeta_2}}
d\hat{k}=\text{const} \ ,\nonumber
\end{eqnarray}
where we have used the fact that in an isotropic 2-dimensional
systen we have $Z^{(2)}(\B k)\sim F^{(2)}(\B k) \propto
k^{-2-\zeta_2}$. We note that the constancy of this integral
should be judged on the background of the decaying function, as
done in  Fig. \ref{fig4-3}. We see that while the 2nd order
structure function decays over three orders of magnitude, the
``constant" objects changes by a factor of 2. The lack of
constancy can be attributed to the sensitivity to the outer scale
as seen in the data collapse in Fig. (\ref{fig4-1}). If the
collapse on the left side of the curve were perfect, so would be
the constancy of the weighted integral. Note that in the
calculation we have employed $\zeta_2 = 0.67$ in agreement with
\cite{00CLMV}
%%%%%%%%%%%%%%%%%%%%%%%%%%%%%
%%%%%%%%%%%%%%%%%%%%%%%%
\subsection{analysis of the multi-point propagators}
%%%%%%%%%%%%%%%%%%%%%%%%%%%%%
\begin{figure}
\centering
\psfrag{yl}{$t^{\frac{\zeta_4}{\zeta_2}} \hat\C P^{(4)}(k,300,t)$}
\psfrag{xl}{$k t^{1/\zeta_2}$}
\includegraphics[width=.45\textwidth]{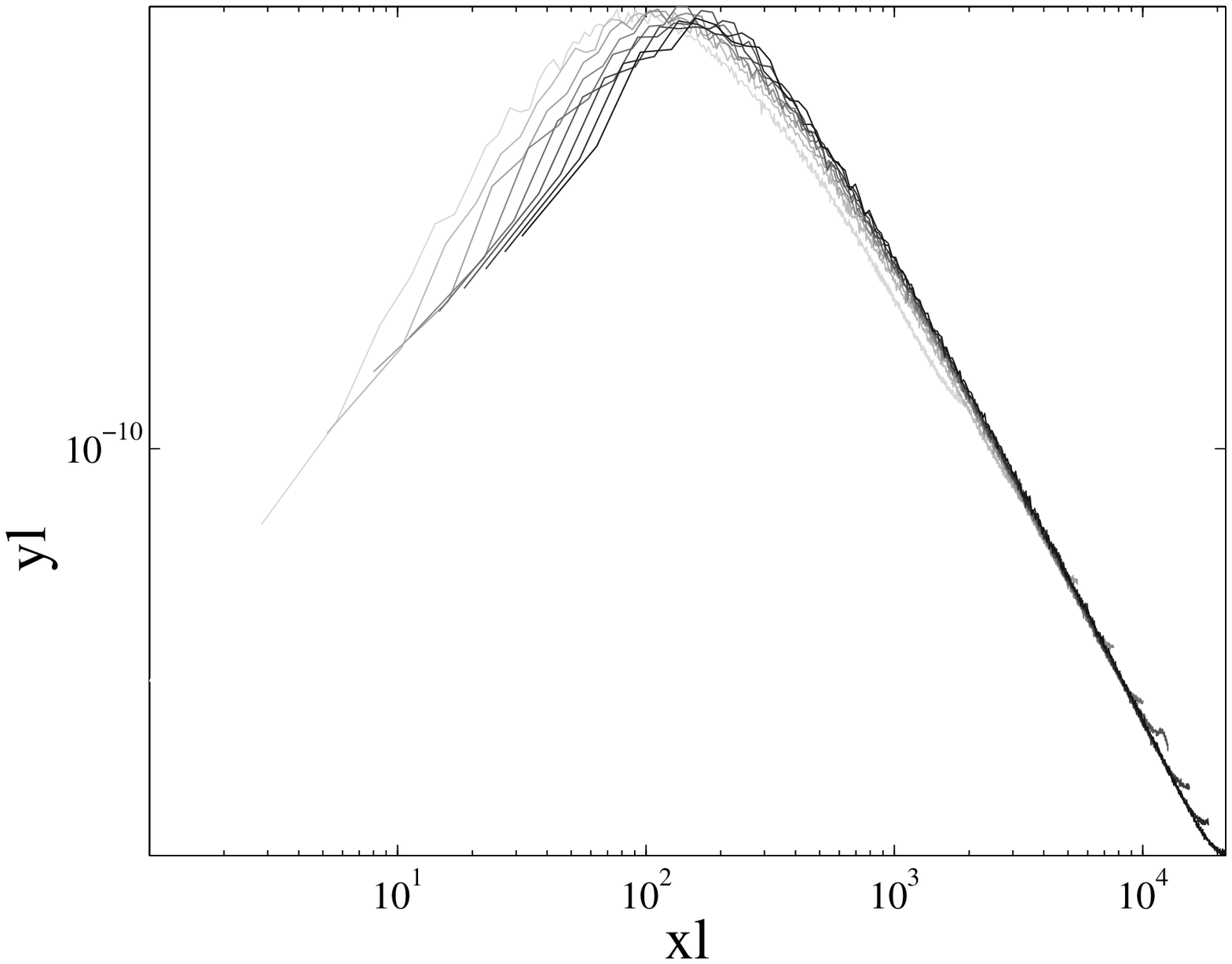}
\psfrag{yl}{$t^{\frac{\zeta_6}{\zeta_2}} \hat\C P^{(6)}(k,300,t)$}
\psfrag{xl}{$k t^{1/\zeta_2}$}
\includegraphics[width=.45\textwidth]{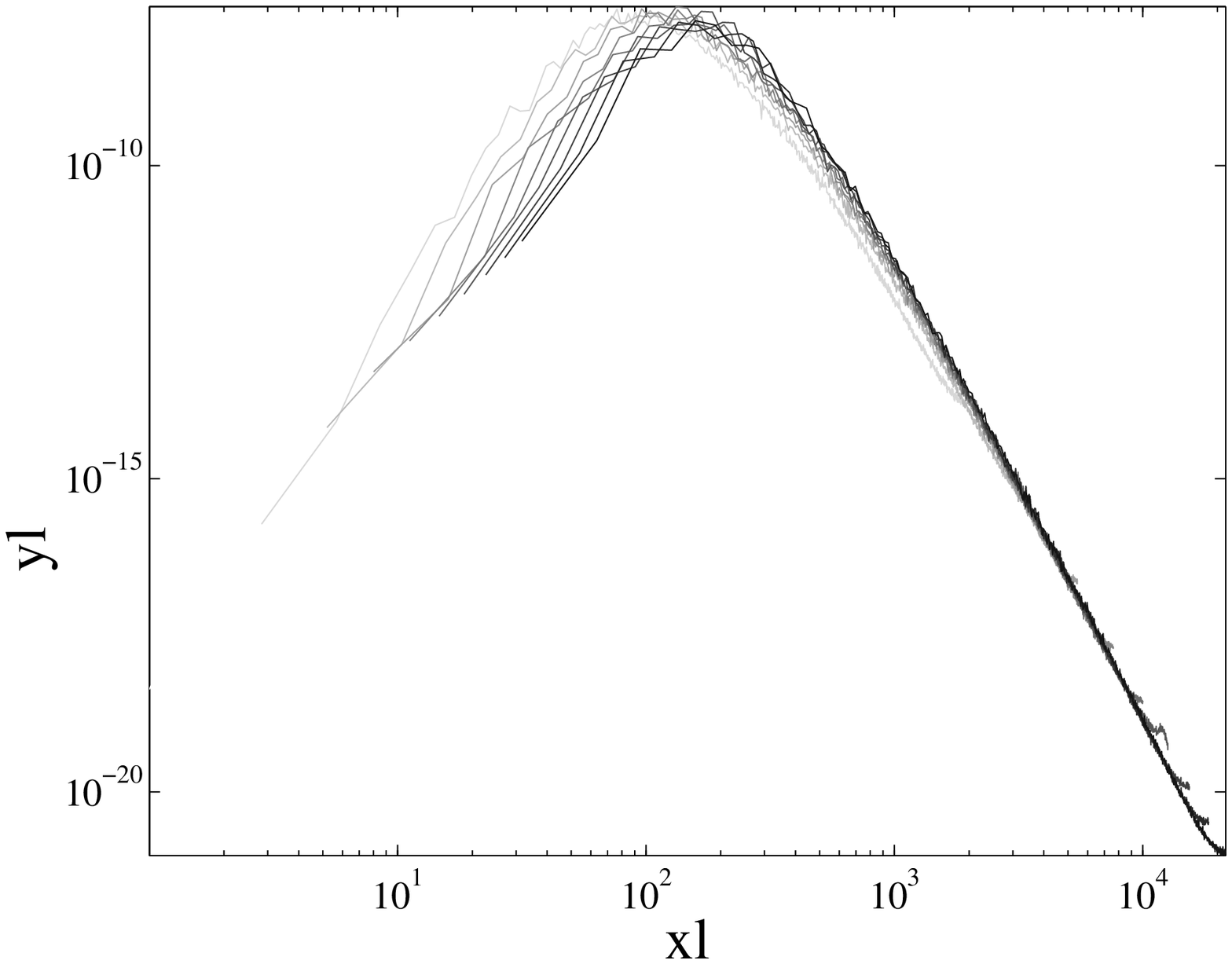}
\caption{$t^{\frac{\zeta_4}{\zeta_2}}\hat \C P^{(4)}(k,300,t)$ (upper panel) and
$t^{\frac{\zeta_6}{\zeta_2}}\hat \C
  P^{(6)}(k,300,t)$ (lower panel) for 10 different times (light grey earliest dark grey last)}
\label{fig4-4}
\end{figure}
%%%%%%%%%%%%%%%%%%%%%%%%
Examining Eq.(\ref{phimult}), we expect that for long times the
functional form of the propagator for the fused correlation
function defined in Eq.(\ref{defpropdmult}) is:
\begin{equation}
\hat \C P^{(2N)}(k,k',t)\propto\frac{1}{t^{\zeta_{2N}/\zeta_2}}
H^{(2N)}(kt^{1/\zeta_2}) \ ,
\end{equation}
where $\zeta_{2N}$ is the leading scaling exponent for the $2N$'th
correlation function. In Fig. \ref{fig4-4} we demonstrate the
data collapse obtained by assuming this form  for the 4'th and
6'th order propagators. We notice the same excellent collapse on the right hand side of
the function as in the 2nd order propagator. Also the problems
with the outer scale show up in a similar manner, giving less
than impressive collapse of the left hand part of the function.
For the present data collapse we employed simple scaling
$\zeta_{2N} = N\zeta_2$; our data does not support strongly
anomalous exponents.

\begin{figure}
\centering
\psfrag{yl}{$\int_0^\infty \hat \C P^{(4)}(k,300,t)dk$}
\psfrag{xl}{$t$}
\includegraphics[width=.45\textwidth]{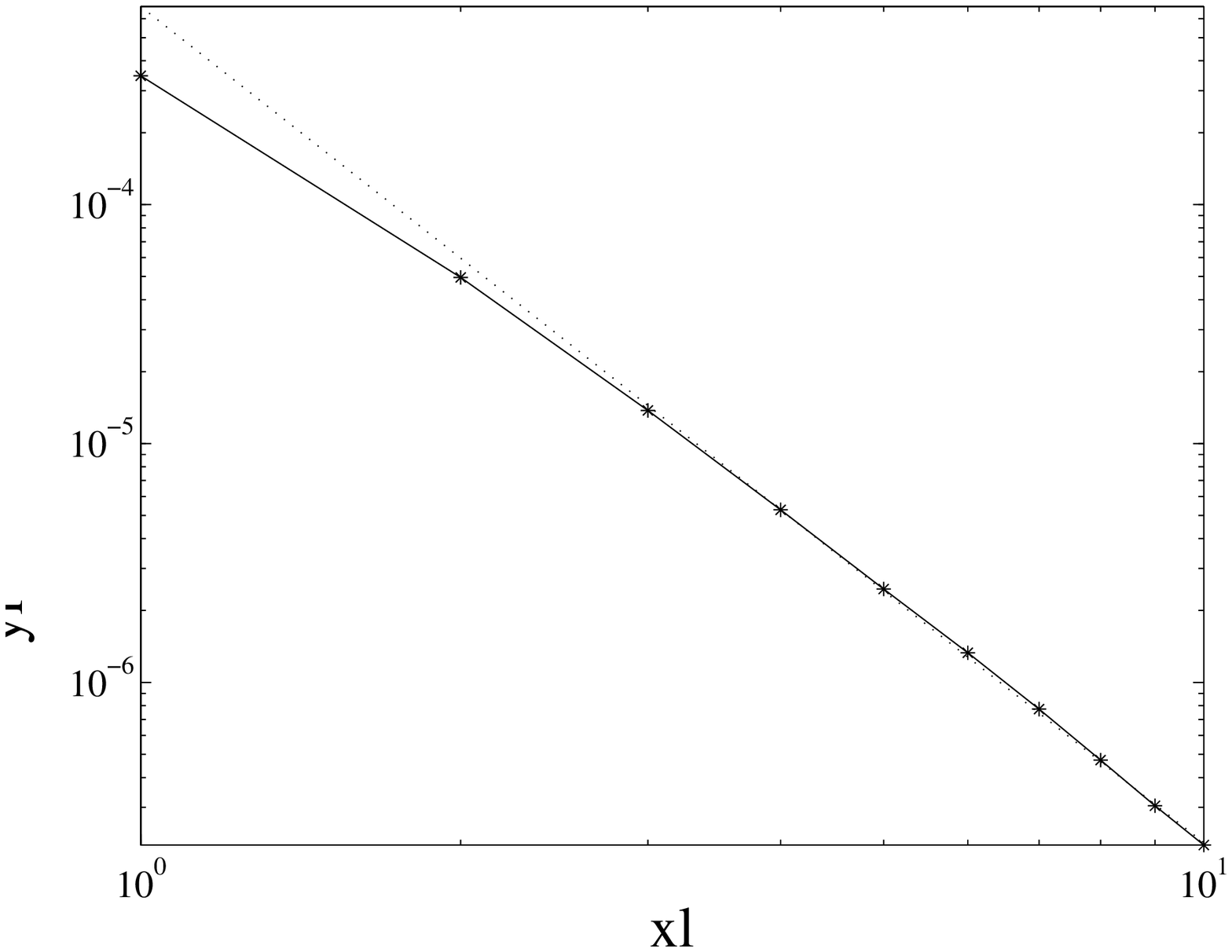}
\psfrag{yl}{$\int_0^\infty \hat \C P^{(6)}(k,300,t)dk$}
\psfrag{xl}{$t$}
\includegraphics[width=.45\textwidth]{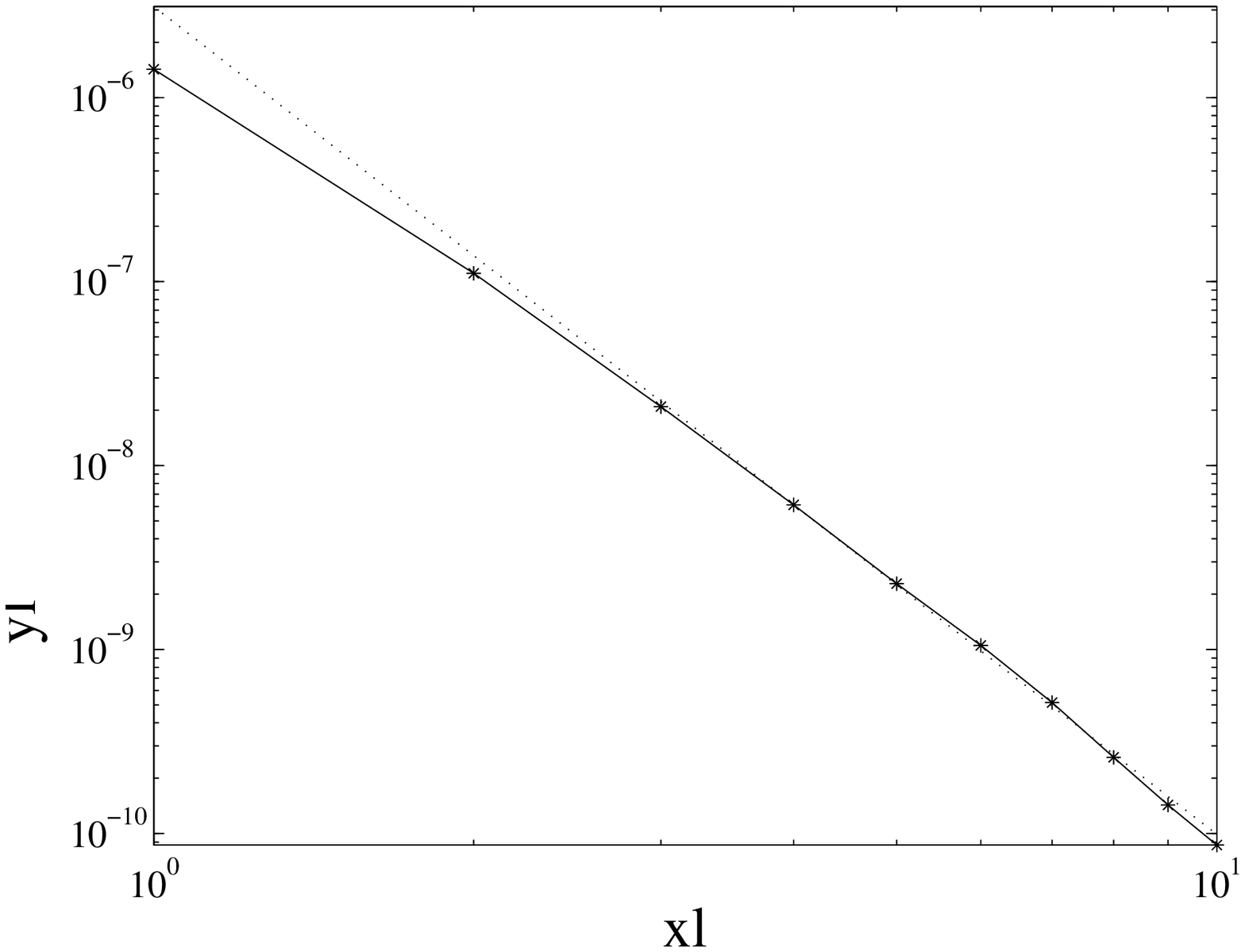}
\caption{The integral of the 4-point (upper panel) and 6-point (lower
  panel) propagators in solid line, $\int_0^\infty \hat \C
  P^{(2m)}(k,300,t)d^2k$, compared with the expected
  $t^{-(1+\zeta_{2N})/\zeta_2}$, in the dotted line. } \label{fig4-5}
\end{figure}

\begin{figure}
\centering
\psfrag{yl}{$\int Z^{(4)}( k)\hat \C P^{(4)}( k,300,t) d^2 \B k$}
\psfrag{xl}{$t$}
\includegraphics[width=.45\textwidth]{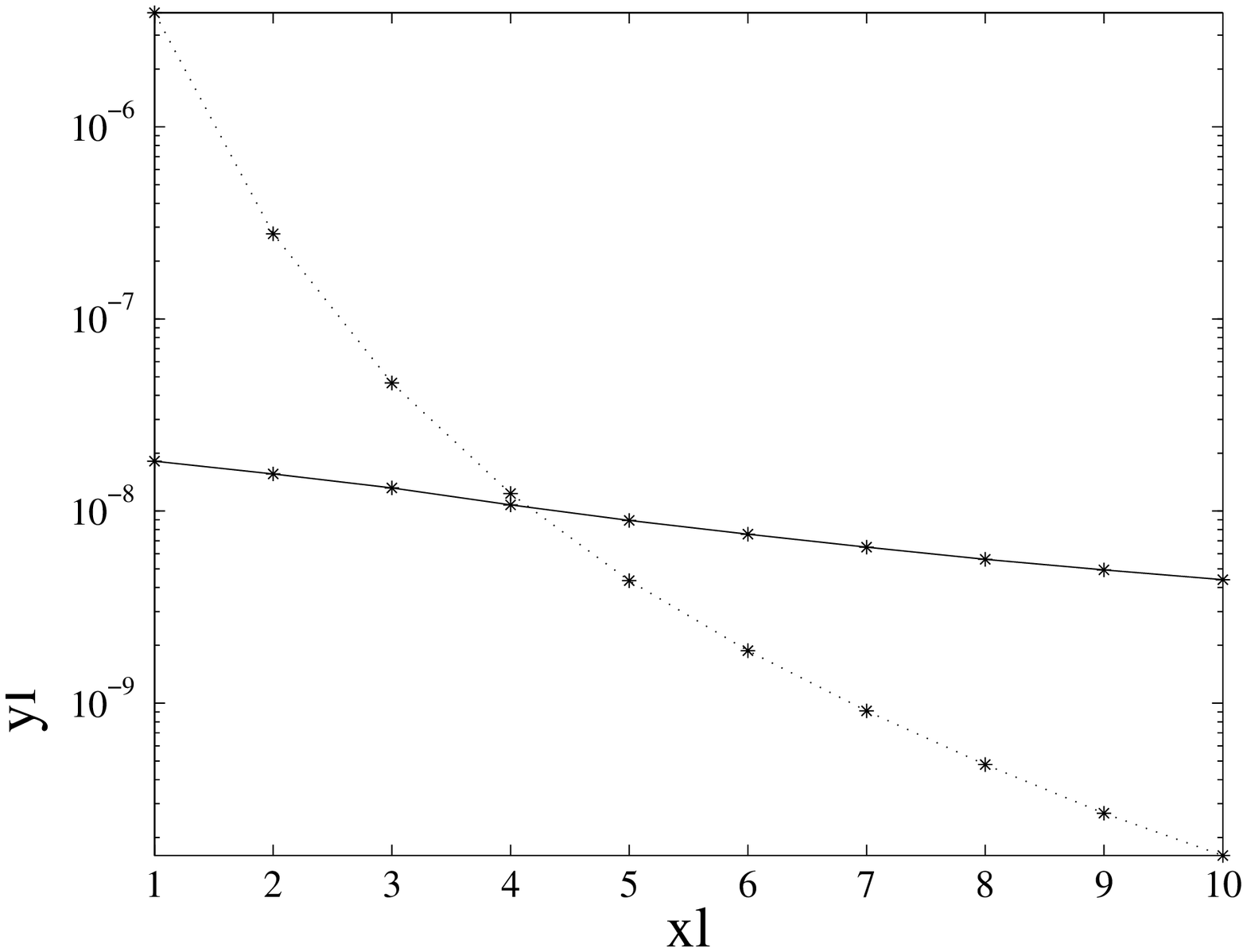}
\psfrag{yl}{$\int Z^{(6)}( k)\hat \C P^{(6)}( k, 300,t) d^2 \B k$}
\psfrag{xl}{$t$}
\includegraphics[width=.45\textwidth]{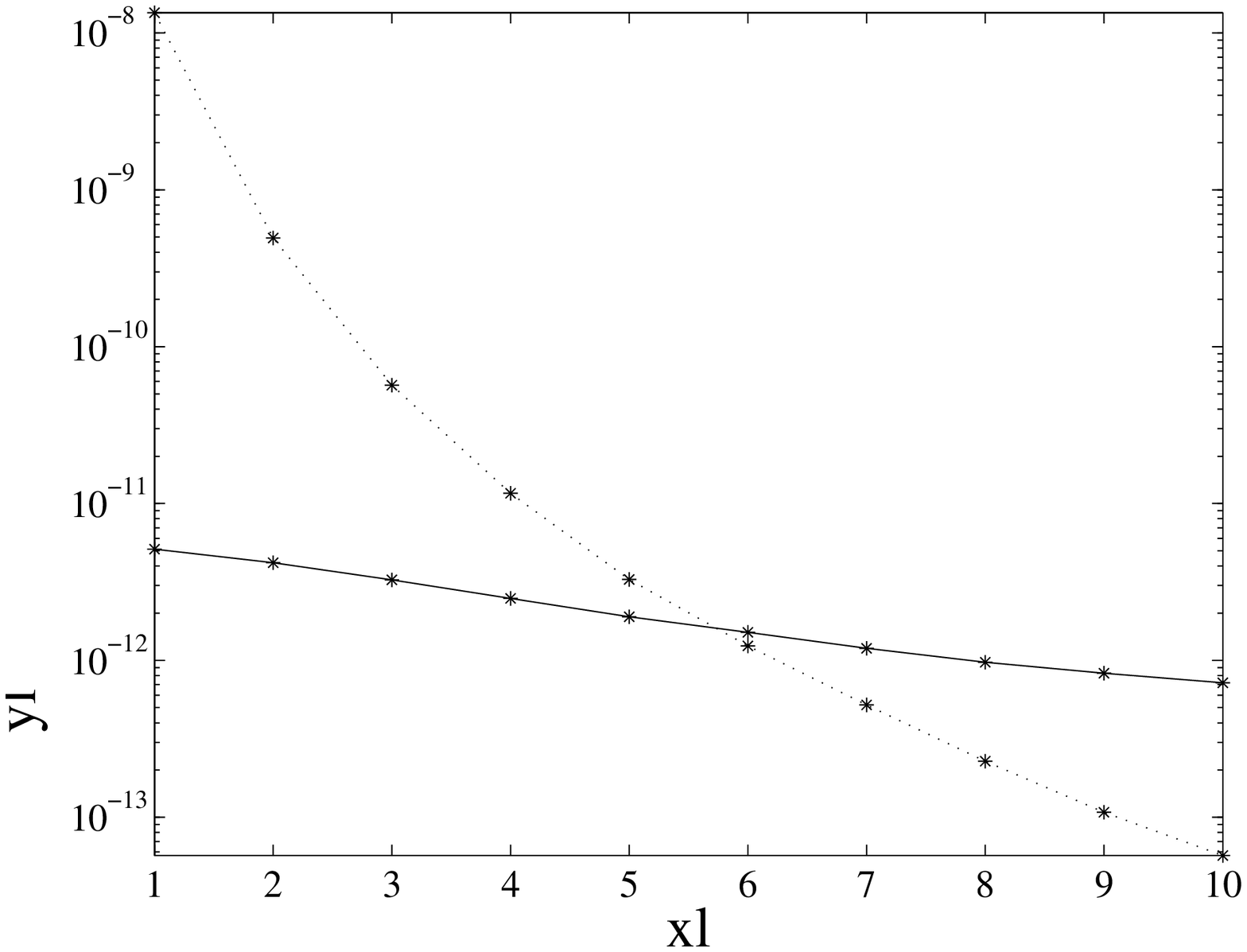}
\caption{The integral of the 4-point propagator $P^{(4)}(k, 300,t)$ (upper
  panel) and the 6-point propagator $P^{(6)}(k,300,t)$ weighed by the left
  zero modes $Z^{(4)}( k)$ and   $Z^{(6)}(k)$ respectively in solid
  line, compared to the integral of the un-weighed object in dotted line}  \label{fig4-6}
\end{figure}

Using the form of the fused multi-point correlators we can again
predict the time behavior of their integral:
\begin{eqnarray}\label{tmult}
\int_0^\infty \hat \C P^{(2N)}(k,k',t)dk &\approx&
\frac{1}{t^{(1+\zeta_{2N})/\zeta_2}}\int_0^\infty  H(\hat
k)d\hat k \nonumber\\
&\propto&\frac{1}{t^{(1+\zeta_{2N})/\zeta_2}}\ .
\end{eqnarray}
In Fig. \ref{fig4-5} the time dependence of the integrals is
plotted, showing agreement of similar quality to Fig.
\ref{fig4-2}.
%%%%%%%%%%%%%%%%%%%%%%%%%%%%%
%%%%%%%%%%%%%%%%%%%%%%%%
As in the 2-point case, weighing the fused multi-point functions
by the appropriate multi point fused steady state correlators
should yield a constant:
\begin{eqnarray}
&&\int  Z^{(2N)}(k)\hat{\C P}^{(2N)}( k,k',t) d^2 \B k
\\&&\propto\int_0^\infty  \frac{H^{(2N)}(kt^{1/\zeta_2})}
{k^{2+\zeta_{2N}} t^{\zeta_{2N}/\zeta_2}}k dk
=\int_0^\infty \frac{H^{(2N)}(\hat{k})}{\hat k^{1+\zeta_{2N}}}
d\hat{k}=\text{const} \ ,\nonumber
\end{eqnarray}
where $Z^{2N}(k)\propto k^{-2-\zeta_{2N}}$. The quality of the
constancy is demonstrated in Fig. \ref{fig4-6}. Again finite size
effects lead to some decrease in time of these weighted objects,
which nevertheless is very much reduced compared to the decaying
correlation functions.
%%%%%%%%%%%%%%%%%%%%%%%%%%%%%
%%%%%%%%%%%%%%%%%%%%%%%%
\section{concluding remarks}
We have demonstrated in this paper that the generic advection of
a passive scalar by a velocity field that obeys the Navier-Stokes
equations can be discussed in terms of Eulerian Statistically
Preserved Structures. By initiating a decay with delta-function
initial conditions (concentrated on $k=300$) we have found
numerically the corresponding columns of the time-dependent
propagators for the 2nd, 4th and 6th order correlation functions
(where for the 4th and 6th order objects we considered partial
(``fused") information). Note that in contrast to the Lagrangian
formulation of Statistically Preserved Structures \cite{01CV},
for which there is no Preserved Structure corresponding to the
2nd order correlation, in the Eulerian formulation such an object
exists and had been analyzed explicitly. We have used a simple
(non-generic) model of passive scalar advection to guess the
analytic scaling form of the propagators in the generic problem.
The test for the relevance of this guessed form is the data
collapse shown in Fig. 5 and 8. The guessed time dependence
appears to be in close correspondence with the data as shown in
Figs. 6 and 9.

The analytic forms of the propagators predict the existence of
eigen-modes with eigenvalue 1 (Statistically Preserved
Structures). We believe that this is the first demonstration of
Eulerian Statistically Preserved Structures in a generic flow. The
numerical evidence for the constancy of the latter is
encouraging, if not fully conclusive, as seen in Figs. 7 and 10.
We attributed the (relatively small) decrease in amplitude of the
putative Statistically Preserved Structures to the less than
perfect data collapse at the largest scales (smallest
$k$-vectors) that are seen in Figs. 5 and 8. These in turn stem
from the intervention of the outer scale in our scaling range, a
boundary effect that we did not succeed to eliminate in our
modest-size simulations. It would be interesting to see whether
larger 2-d simulations could remove this finite-size effect to
demonstrate conclusively the constancy of the Statistically
Preserved Structures.

In concluding we wish to point out an additional benefit to the
present formulation. Usually in modeling turbulent advection it
is customary to resort to dubious concepts such as `turbulent
diffusion' in order to write a diffusion equation for the
correlation functions. The present approach indicates a much
better procedure, i.e. to find, for a given turbulent field, the
form of the propagator which can be then used to provide analytic
predictions for any initial (t=0) correlation function. This
procedure may be quite attractive for more complex hydrodynamic
flows with unusual boundaries or coherent structures. We believe
that when the turbulence is sufficiently well developed scaling
forms for the propagator will exist, and once found can be used
to solve efficiently any statistical initial value problem.
%%%%%%%%%%%%%%%%%%%%%%%%%%%%%%%%%%%%%%%%%%%%%%%%%%%
\begin{acknowledgments}
We thank Antonio Celani for providing us with his code for DNS of
2-dimensional turbulent advection. This work had been supported
in part by the European Commission under a TMR grant, the
Minerva Foundation, Munich, Germany, and the Naftali and Anna
Backenroth-Bronicki Fund for Research in Chaos and Complexity.

\end{acknowledgments}


\begin{thebibliography}{99}
\bibitem{79MY}
A.S. Monin and A.M. Yaglom, {\em Statistical Fluid Mechanics}, vol. 1, chapter 5,
(MIT, Cambridge 1979).

\bibitem{98BGK}
D. Bernard, K. Gawedzki, A. Kupiainen, J.Stat. Phys {\bf 90}, 519 (1998)

\bibitem{01CV}
A. Celani and M. Vergassola, Phys. Rev. Lett. {\bf 86}, 424
(2001).

\bibitem{01ABCPV}
I. Arad, L. Biferale, A. Celani, I. Procaccia and M. Vergassola,
Phys. Rev. Lett. {\bf 87}, 164502 (2001).

\bibitem{02CGP}
Y. Cohen, T. Gilbert and I. Procaccia, Phys. Rev. E, {\bf 65},
026314 (2002).

\bibitem{97BBW}
R. Benzi, L. Biferale and A. Wirth, Phys. Rev. Lett. {\bf 78}, 4926 (1997)

\bibitem{00CLMV}
A. Celani,  A. Lanotte, A. Mazzino and M. Vergassola, Phys. Rev. Lett. {\bf 84}, 2385
(2000).

\end{thebibliography}
\end{document}